# Nonlinearity management of photonic composites and observation of spatial-modulation instability due to quintic nonlinearity

Albert S. Reyna* and Cid B. de Araújo

*Departamento de Física, Universidade Federal de Pernambuco, 50670-901, Recife, PE, Brazil*

*Corresponding author. E-mail: areynao@yahoo.com.br

**Abstract**

We present a procedure for nonlinearity management of metal-dielectric composites. Varying the volume fraction occupied by silver nanoparticles suspended in acetone we could cancel the refractive index related to the third-order susceptibility, $\chi_{eff}^{(3)}$, and the nonlinear refraction behavior was due to the fifth-order susceptibility, $\chi_{eff}^{(5)}$. Hence, in a cross-phase modulation experiment, we demonstrated for the first time the effect of spatial-modulation-instability due to $\chi_{eff}^{(5)}$. The results are corroborated with numerical calculations based on a generalized Maxwell-Garnet model.

The nonlinear (NL) response of matter to optical fields can be described expressing the induced polarization by a power series of the field with NL susceptibilities, $\chi^{(N)}$, $N = 2, 3, \ldots$, as coefficients of the series [1]. Since all even-order susceptibilities are null in systems with inversion symmetry, the lowest order NL response is generally due to the third-order susceptibility, $\chi^{(3)}$, which contributes for generation of fields that depend on the cubic power of the incident field. Therefore, most of the NL studies are related to $\chi^{(3)}$ that is responsible for two-photon absorption, third-harmonic generation, and coherent Raman scattering, among other effects. Cascade processes of $\chi^{(3)}$ behave analogously to high-order nonlinearities (HON) and were reported for gases and condensed matter systems [2]. Nevertheless, experiments based on HON, related to direct (not cascade) processes, have been reported for a large variety of systems [3].

HON are still under investigation from the fundamental point of view [4] and there is large interest in phenomena such as liquid light condensates [5], soliton formation [6, 7] and other transverse NL effects [7, 8]. Interferences between third- and fifth-order processes have been reported for different systems [9]. It is also of interest the exploitation of HON in quantum information [10], quantum memories [11] and for improvement of high-precision measurements [12].

The interest in novel effects related to quintic and cubic-quintic nonlinearities lead several authors to propose experiments with metal-dielectric nanocomposites (MDNC) [7]. From the basic point of view MDNC are interesting systems because their NL response can be controlled by changing the nanoparticles (NPs) volume fraction, $f$ (the ratio between the volume occupied by the NPs and the host). Indeed, the interest in the NL properties of MDNC is large in nanoscience and nanotechnology [13-18].

In this paper we present a procedure for nonlinearity management of a MDNC aiming its exploitation for practical realization of mathematical models and experiments related to HON [19]. The NL response of a colloid consisting of silver NPs suspended in acetone was described by effective susceptibilities, $\chi_{eff}^{(2N+1)}$, $N = 1, 2, 3, \ldots$, that depend on the NL susceptibilities of the host liquid and the NPs. HON up to seventh-order were measured as a function of the NPs volume fraction. In particular for $f = 1.6 \times 10^{-5}$ we obtained $\mathrm{Re}\,\chi_{eff}^{(3)} = 0$ and $\mathrm{Re}\,\chi_{eff}^{(5)} \neq 0$. Using this condition two-beam cross-phase modulation experiments were performed and revealed for the first time the effect of spatial-modulation-instability due to the fifth-order susceptibility.

The silver colloid was prepared as described in [13, 20]: 90 mg of AgNO$_3$ were diluted in 500 ml of water at 100°C; 10 ml of solution of 1% sodium citrate was added for reducing the Ag$^+$ ions, and later was boiled and strongly stirred for 1 h. Subsequently, photo-fragmentation of the NPs [21] was performed by placing a cuvette containing the pristine colloid in front of the second harmonic beam from a Q-switched Nd: YAG laser (10 Hz, 8 ns, 85 mJ/pulse) for 1 h while the suspension was slowly stirred. A homogeneous distribution of spherical NPs (average diameter: $9.0 \pm 2.2$ nm) was obtained. Hence, colloids prepared by adding 20 μL to 300 μL of the Ag-water suspension in 1 mL of acetone, with $f$ varying from $0.5 \times 10^{-5}$ to $7.5 \times 10^{-5}$, were used in the experiments.

The linear absorption spectra of the samples were measured with a commercial spectrophotometer. For the NL measurements we used the second harmonic of a Q-switched and mode-locked Nd: YAG laser (80 ps, 532 nm); single pulses at 10 Hz were selected using a pulse-picker. The incident pulses on the samples had maximum energy of 10 μJ.

Figure 1 shows the absorption spectra of the colloid before (pristine) and after photo-fragmentation of the NPs using 8 ns pulses with 85 mJ/each. This procedure allows obtaining a narrow size distribution of NPs as demonstrated in [13, 21] and by the narrow spectrum of Fig.1. In addition, the spectrum of pure acetone is shown to assure that the band at $\approx 400$ nm (linewidth $\approx 50$ nm) is due to the surface plasmon resonance associated to the NPs.

The Z-scan technique [22] was used for characterization of the samples. For the measurements the sample was moved along the beam propagation direction (Z axis) in the region where the laser beam is focused. Measurements of the transmitted beam intensity through a small aperture placed in front of a photodetector in the far-field region provide the NL refractive index value (*closed-aperture* scheme). When all light transmitted through the sample is detected, the NL absorption coefficient can be determined (*open-aperture* scheme). For both schemes the laser beam was focused by a 5 cm focal distance lens, producing a beam waist of 20 µm at the focus. The detected signals were processed by boxcar integrators and computer. To improve the signal-to-noise ratio a reference channel was used as in [23]. Carbon disulfide ($CS_2$) with NL refractive index equal to $+3.1 \times 10^{-14}$ cm$^2$ / W [22] was the reference standard for calibration of the measurements.

Figure 2 shows *closed-aperture* Z-scan traces corresponding to four $f$ values. The colloid inside a 1 mm long quartz cell was scanned along the Z-axis using a translation stage. Figures 2(a) and 2(b) show profiles that indicate positive NL refractive index for $f = 0.8 \times 10^{-5}$ and $1.3 \times 10^{-5}$, respectively. The normalized peak-to-valley transmittance change, $\left| \Delta T_{PV} \right|$, is smaller in Fig. 2(b) than in Fig. 2(a) because the NPs contribute to the NL refractive index with opposite sign than acetone that has $n_2 = +2.16 \times 10^{-15}$ cm$^2$ / W [24]. Figures 2(c) and 2(d) corresponding to $f \geq 2 \times 10^{-5}$ indicate that the NL refractive index of the colloid became negative because the silver NPs dominate the NL response. For small laser intensity ($I = 2.0\,\mathrm{GW/cm^2}$) we determined $n_2$ for different $f$ values using the expression $\left|\Delta T_{PV}\right| = 0.406\,k\,L_{eff}\,n_2\,I$ from ref. [22], where $k = 2\pi n_0/\lambda$, $L_{eff} = \left[1 - \exp(-\alpha_0 L)\right]/\alpha_0$, $L$ is the sample length, $\alpha_0$ is the linear absorption coefficient and $\lambda$ is the laser wavelength. The sign-reversal of $n_2$ as a function of $f$ was observed for all intensities used. However no sign-reversal of $n_2$ was observed fixing $f$ and changing $I$ for the whole range of $f$ values.

Figure 3(a) shows new features in the *closed-aperture* Z-scan profiles due to HON for $I > 6.0\,\mathrm{GW/cm^2}$ and Fig. 3(b) shows *open-aperture* profiles for various laser intensities. The solid curves were obtained following the procedure of [13]. After each Z-scan experiment no changes were observed in the linear absorption spectrum indicating that the energy of the laser pulses used is not enough to modify the NPs. The NL experiments were repeated more than one time with each sample and the results were reproduced.

To analyze the results we plotted $\left|\Delta T_{PV}\right|/I$ versus $I$. For $f < 0.8 \times 10^{-5}$ the ratio $\left|\Delta T_{PV}\right|/I$ remains constant for intensities up to $10\,\mathrm{GW/cm^2}$, indicating negligible contributions of $\chi_{eff}^{(2N+1)}$ ($N > 1$). For $f \approx 1.3 \times 10^{-5}$, the ratio $\left|\Delta T_{PV}\right|/I$ presents linear dependence with $I$ and from the slope of the straight line we determined $n_4 \propto \mathrm{Re}\,\chi_{eff}^{(5)}$ [13]. For $f \geq 2.0 \times 10^{-5}$ the intensity dependence of $\left|\Delta T_{PV}\right|/I$ is a polynomial function that allows obtaining the refractive indices associated to NL susceptibilities up to the seventh-

order. For instance, when $f = 2.5 \times 10^{-5}$ and $I = 9.0\,\mathrm{GW/cm^2}$ we determined $n_2 = -1.1 \times 10^{-15}\,\mathrm{cm^2/W}$, $n_4 = +6.9 \times 10^{-25}\,\mathrm{cm^4/W^2}$, and $n_6 = -1.1 \times 10^{-34}\,\mathrm{cm^6/W^3}$.

$\left|\Delta T_{PV}\right|/I$ also exhibited a polynomial dependence with $I$ in the *open-aperture* experiments and from the results in Fig 3(b) we obtained $\alpha_i$ ($i = 2, 4, 6$) for different values of $f$. For example, for $f = 5.0 \times 10^{-5}$, we have $\alpha_2 = -4.9 \times 10^{-10}\,\mathrm{cm/W}$, $\alpha_4 = +1.4 \times 10^{-19}\,\mathrm{cm^3/W^2}$ and $\alpha_6 = -1.7 \times 10^{-29}\,\mathrm{cm^5/W^3}$.

Figure 4 summarizes the results of $n_2$, $n_4$, $\alpha_2$, and $\alpha_4$ when $I = 9.0\,\mathrm{GW/cm^2}$. All parameters show linear dependence with $f$. We remark that for $f \approx 1.6 \times 10^{-5}$ we have $n_2 = 0$ but $n_4 = +3.2 \times 10^{-25}\,\mathrm{cm^4/W^2}$. This result does not violate the powers series of the NL polarization and opens new routes for exploitation of novel effects considering that under the conditions identified here $\mathrm{Re}\,\chi_{eff}^{(5)}$ is the lowest order NL refractive response.

Therefore, spatial cross-phase modulation experiments were performed to exploit the response of the MDNC with adjustable $f$ values. The laser beam was split in probe and pump beams with intensity ratio 1:10. The probe beam was weak enough to not induce self-focusing. The two beams were aligned to counter-propagate along the sample and a careful adjustment of the spatial and temporal overlap between the pulses was made. The measured beam waists were $\approx 100$ μm with Rayleigh lengths of $\approx 6$ μm. The beams' profiles were analyzed using a CCD camera.

Figures 5(a)–5(c) shows the probe beam profiles when the pump beam with $2.0\,\mathrm{GW/cm^2}$ is present; the results correspond to $f = 0.5 \times 10^{-5}$, $1.6 \times 10^{-5}$ and $2.5 \times 10^{-5}$, respectively. We analyzed the spatial profile of the probe beam as an intensity matrix; the curves in Fig. 5(d)–5(f) represent column matrix components passing through the axis of the probe beam exhibiting the beam intensity versus the radial coordinate. As expected, it was observed spatial broadening of the probe beam due to the presence of the pump beam. Moreover, new spatial frequencies were generated due to the spatial-modulation-instability (SMI) induced by the pump beam as noticed by the feature induced in the center of the probe beam profile. Due to the $f$ values used, Fig. 5(d) and Fig. 5(e) exhibit the SMI effect due to only $\chi_{eff}^{(3)}$ and only $\chi_{eff}^{(5)}$, respectively; Fig. 5(f) illustrates the simultaneous influence of $\chi_{eff}^{(3)}$ and $\chi_{eff}^{(5)}$. Figures 5(g)-5(i) show the numerical results obtained as discussed below.

To understand the results we developed an extension of the Maxwell-Garnet model [25] including the contributions of the third- and fifth-order susceptibilities. The quasi-static approximation was assumed because the NPs diameters are smaller than $\lambda$. Accordingly, for $f \ll 1$ and optical electric field, $\boldsymbol{E}_0$, the induced polarization can be written as $\boldsymbol{P} = \boldsymbol{P_h} + \dfrac{1}{V}\sum_{i=1}^{N_p}\boldsymbol{p_i}$, where $\boldsymbol{P_h}$ is the host polarization, $N_p$ is the number of NPs inside the volume $V$ and $\boldsymbol{p_i} = \varepsilon_h\sigma_i\boldsymbol{E}_0$ is the induced dipole moment of each NP; $\sigma_i$ is the NP polarizability given by $\sigma_i = 3v_i\left(\dfrac{\varepsilon_{np} - \varepsilon_h}{\varepsilon_{np} + 2\varepsilon_h}\right)$, where $\varepsilon_{np}$ ($\varepsilon_h$) is the dielectric function of the NPs (host) and $v_i$ is the volume of the NP. The dielectric functions can be expressed as a sum of the linear and NL contributions such that $\varepsilon_{h,np} = \varepsilon_{h,np}^{(\mathrm{L})} + \varepsilon_{h,np}^{(\mathrm{NL})}$, where the NL terms are given by $\varepsilon_{np}^{(NL)} = \dfrac{3}{4}\chi_{np}^{(3)}\left\langle\left|\boldsymbol{E}_{np}\right|^2\right\rangle + \dfrac{5}{8}\chi_{np}^{(5)}\left(\left\langle\left|\boldsymbol{E}_{np}\right|^2\right\rangle\right)^2$ and $\varepsilon_h^{(NL)} = \dfrac{3}{4}\chi_h^{(3)}\left\langle\left|\boldsymbol{E}_0\right|^2\right\rangle$, with $\chi_{np}^{(i)}$, $i = 3, 5$, and $\chi_h^{(3)}$ being the $i$-th NL susceptibility of the NPs and the third-order

susceptibility of the host, respectively. The expression $\left\langle \left| \boldsymbol{E}_{np} \right|^2 \right\rangle = \left| \eta \right|^2 \left| \boldsymbol{E}_0 \right|^2$ represents the mean squared modulus of the electric field inside the NP and $\eta = 3\varepsilon_h / (\varepsilon_{np} + 2\varepsilon_h)$ is the local field factor. The effective NL susceptibilities are determined expanding the polarizability up to second order in $\left| \boldsymbol{E}_0 \right|^2$ to obtain

$$\chi_{eff}^{(3)} = f\, L^2 \left| L \right|^2 \chi_{np}^{(3)} + \chi_h^{(3)}, \tag{1}$$

$$\chi_{eff}^{(5)} = f\, L^2 \left| L \right|^4 \chi_{np}^{(5)} - \frac{6}{10} f\, L^3 \left| L \right|^4 \left( \chi_{np}^{(3)} \right)^2 - \frac{3}{10} f\, L \left| L \right|^6 \left| \chi_{np}^{(3)} \right|^2, \tag{2}$$

where $L = 3\varepsilon_h^{(L)} / (\varepsilon_{np}^{(L)} + 2\varepsilon_h^{(L)})$.

Equation (1) corroborates our interpretation of Fig. 2, i.e.: the sign-reversal of $\mathrm{Re}\, \chi_{eff}^{(3)}$ as a function of $f$ is due to competition between the terms containing $\mathrm{Re}\, \chi_{np}^{(3)}$ and $\mathrm{Re}\, \chi_h^{(3)}$ that have opposite signs. Equation (2) was derived considering negligible terms proportional to $\left( \chi_h^{(3)} \right)^2$ and $\left( \chi_h^{(3)} \chi_{np}^{(3)} \right)$ due to the magnitude of $\chi_h^{(3)} = +1.67 \times 10^{-21} \left( \mathrm{m}^2 / \mathrm{V}^2 \right)$ that is $\approx 5$ order of magnitude smaller than $\chi_{np}^{(3)}$. The fifth-order susceptibility of the host was neglected in comparison with the contributions due to the NPs. It can be seen from Eq. (2) that the NPs contributions to the direct fifth-order susceptibility, $\chi_{np}^{(5)}$, and cascade of third-order susceptibility (the terms with $\chi_{np}^{(3)}$) are enhanced due to the high powers of $L$ and $\left| L \right|$. The NL response of the NPs is attributed mainly to the $s$-electrons in the conduction band because the energy difference between the $d$-band and the Fermi level is $\approx 4\,\mathrm{eV}$, larger than the energy of the incident photons ( $2.34\,\mathrm{eV}$ ).

From Eqs. (1) and (2), we verified that $\mathrm{Im}\,\chi_{eff}^{(3)}$ ($\mathrm{Re}\,\chi_{eff}^{(5)}$) is negative (positive) for all values of $f$, in agreement with the experiments. Considering $\varepsilon_{np}^{(L)} = (0.055 + i3.455)^2$ [26] and $\varepsilon_h^{(L)} = (n_0)^2$, with $n_0 = 1.36$, we obtained $\chi_{np}^{(3)} = -5.9 \times 10^{-16} - i8.5 \times 10^{-16} \left(\mathrm{m}^2/\mathrm{V}^2\right)$, $\chi_{np}^{(5)} = -1.0 \times 10^{-33} - i1.7 \times 10^{-31} \left(\mathrm{m}^4/\mathrm{V}^4\right)$ and $\chi_{eff}^{(5)} = +3.7 \times 10^{-38} + i3.3 \times 10^{-37} \left(\mathrm{m}^4/\mathrm{V}^4\right)$. For $f = 1.6 \times 10^{-5}$ we have $n_2 \propto \mathrm{Re}\,\chi_{eff}^{(3)} = 0$ and $n_4 \propto \mathrm{Re}\,\chi_{eff}^{(5)} \neq 0$ due to the cascade process associated to $\left(\chi_{np}^{(3)}\right)^2$ and $\left|\chi_{np}^{(3)}\right|^2$, and to the direct process described by $\chi_{np}^{(5)}$.

The cross-phase modulation experiment can be described by coupled equations that relate the pump and probe beams as:

$$-\frac{\partial A_1}{\partial z} - \frac{i}{2k}\left(\frac{\partial^2 A_1}{\partial x^2} + \frac{\partial^2 A_1}{\partial y^2}\right) = \frac{ikn_2}{n_0}\left(|A_1|^2 + 2|A_2|^2\right)A_1 + \frac{ikn_4}{n_0}\left(|A_1|^4 + 6|A_1|^2|A_2|^2 + 3|A_2|^4\right)A_1,$$

$$(3)$$

$$\frac{\partial A_2}{\partial z} - \frac{i}{2k}\left(\frac{\partial^2 A_2}{\partial x^2} + \frac{\partial^2 A_2}{\partial y^2}\right) = \frac{ikn_2}{n_0}\left(|A_2|^2 + 2|A_1|^2\right)A_2 + \frac{ikn_4}{n_0}\left(|A_2|^4 + 6|A_1|^2|A_2|^2 + 3|A_1|^4\right)A_2,$$

$$(4)$$

where $A_1$ and $A_2$ are the slowly varying envelope amplitude of the pump and probe beams, respectively, $k = 2\pi n_0/\lambda$, $x$ and $y$ are the transverse coordinates.

Equations (3) and (4) were solved using the split-step method and considering only one transverse dimension due to the cylindrical symmetry of the beams. The results, considering the values of $n_2$ and $n_4$ determined in the Z-scan experiments, are shown in Fig. 5(g)-5(i) where a good agreement with the experimental results is observed. We emphasize that the results shown in Figs. 5(b), 5(e) and 5(h) represent the first demonstration of SMI due to the fifth-order susceptibility in a system with $n_2 = 0$ and $n_4 \neq 0$.

In summary, we demonstrated the conditions under which the real part of the effective third-order susceptibility of a MDNC can be suppressed and the NL refractive response is dominated by the fifth-order nonlinearity. The experiments demonstrate a procedure for managing cubic-quintic composites in order to develop self-focusing systems with dominant quintic refractive nonlinearity and negligible cubic refraction. As an application we demonstrated the spatial-modulation-instability due to the fifth-order NL refraction for the first time. From the basic point of view such engineered MDNC offer the possibility for practical realization of mathematical models describing the behavior of spatial solitons influenced by HON and other experiments already proposed in the literature. Of course, because of the frequency dependent mismatch between the dielectric functions of the NPs and the host, as well as the excitation of localized surface plasmons in the NPs, the NL susceptibilities of the nanocomposites may be further enhanced by using other laser frequencies.

**Acknowledgments:** We acknowledge financial support from the Brazilian agencies Conselho Nacional de Desenvolvimento Cientifico e Tecnológico (CNPq) and the Fundação de Amparo à Ciência e Tecnologia do Estado de Pernambuco (FACEPE). The work was performed in the framework of the National Institute of Photonics (INCT de Fotônica) project and PRONEX/CNPq/FACEPE.

**Figure captions**

1.  (color online) Linear absorption spectra of acetone (dashed line) and silver colloid before (dotted line) and after photo-fragmentation of the NPs (solid line). Cell length: 1 mm.

2.  (color online) Typical closed-aperture Z-scan traces obtained at 532 nm for different volume fraction $f$. Negative values of $Z$ correspond to locations of the sample between the focusing lens and its focal plane. The solid lines are guides to the eyes. Laser peak intensity: $5.0\,\mathrm{GW/cm^2}$.

3.  (color online) Normalized Z-scan traces obtained for different laser peak intensities; (a) closed-aperture scheme and (b) open-aperture scheme. From bottom to top, the curves correspond to $2\,\mathrm{GW/cm^2}$, $4\,\mathrm{GW/cm^2}$, $6\,\mathrm{GW/cm^2}$, $8\,\mathrm{GW/cm^2}$, and $9\,\mathrm{GW/cm^2}$. Volume fraction: $f = 5.0{\times}10^{-5}$. The curves were normalized and shifted in the vertical to prevent overlap.

4.  (color online) Dependence of the effective third-order coefficients ($n_2 \propto \mathrm{Re}\,\chi_{eff}^{(3)}$ and $\alpha_2 \propto \mathrm{Im}\,\chi_{eff}^{(3)}$) and the effective fifth-order coefficients ($n_4 \propto \mathrm{Re}\,\chi_{eff}^{(5)}$ and $\alpha_4 \propto \mathrm{Im}\,\chi_{eff}^{(5)}$) versus the NPs volume fraction, $f$: (a) $n_2$ and $n_4 I$; (b) $\alpha_2$ and $\alpha_4 I$. Laser peak intensity: $I = 9.0\,\mathrm{GW/cm^2}$.

5.  (color online) Probe beam profile in the presence of the counter-propagating pump beam: (a) $f_1 = 0.5{\times}10^{-5}$; (b) $f_2 = 1.6{\times}10^{-5}$; (c) $f_3 = 2.5{\times}10^{-5}$. Probe beam intensity versus the radial coordinate: (d) $f_1$; (e) $f_2$; (f) $f_3$. Numerical results using the values of $n_2$ and $n_4$ determined in the Z-scan experiments and $f$ as in (d)-(f). Pump beam intensity $2.0\,\mathrm{GW/cm^2}$; probe beam intensity: $0.2\,\mathrm{GW/cm^2}$.

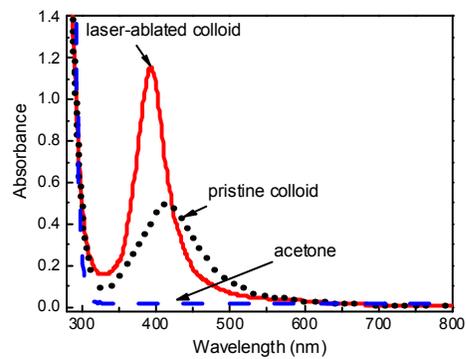

Fig. 1 Reyna et al.

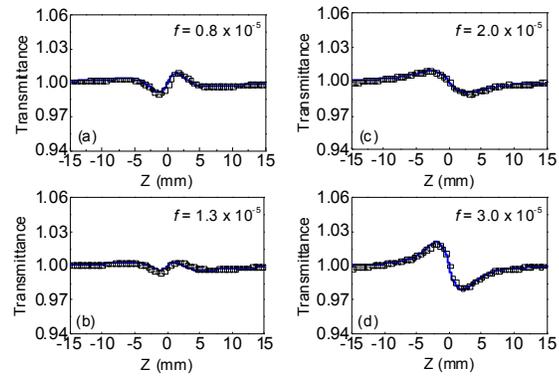

Fig. 2 Reyna et al.

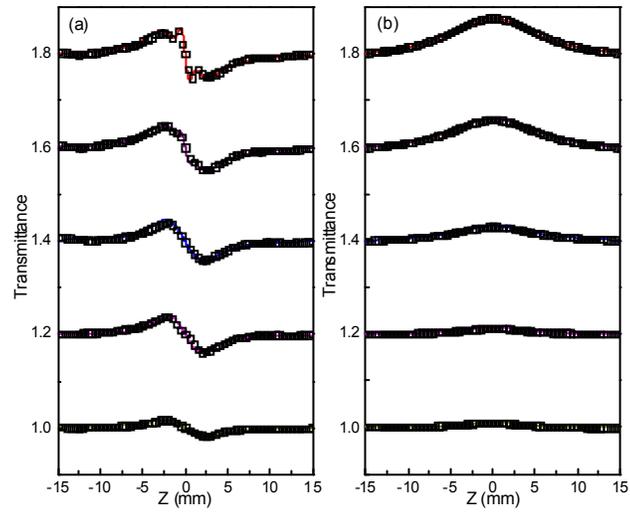

Fig. 3 Reyna et al.

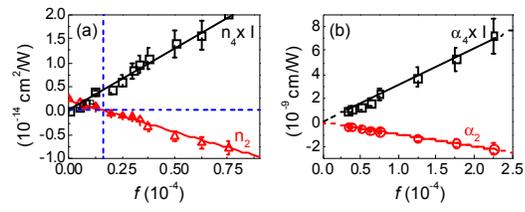

Fig. 4 Reyna et al.

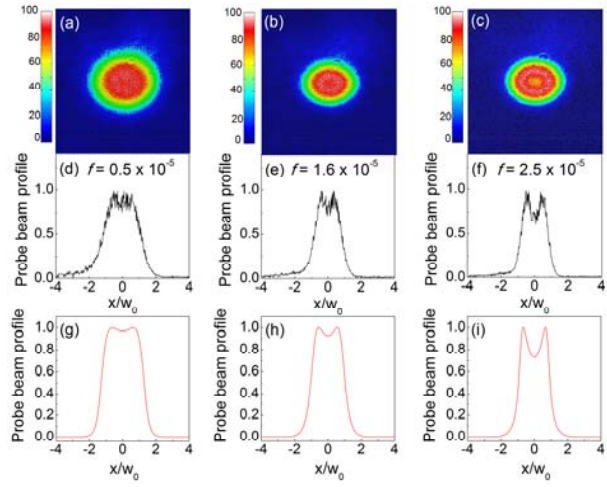

Fig. 5 Reyna et al.